\documentclass{article}
\usepackage{spconf,amsmath,graphicx}
\usepackage[utf8]{inputenc} 
\usepackage[T1]{fontenc}    
\usepackage{hyperref}       
\usepackage{url}            
\usepackage{booktabs}       
\usepackage{amsfonts}       
\usepackage{nicefrac}       
\usepackage{microtype}      
\usepackage{graphicx}
\usepackage{amsmath,amssymb,amsthm}
\usepackage{breqn}
\usepackage[ruled,vlined]{algorithm2e}
\usepackage[utf8]{inputenc}
\usepackage[english]{babel}
\newtheorem{theorem}{Theorem}
\graphicspath{ {./images/} } 
\usepackage{setspace}
\usepackage{mathtools}
\usepackage[title]{appendix}
\usepackage{color}
\usepackage{multirow}
\usepackage{tabularx}
\newtheorem{proposition}{Proposition}

\ninept
\title{Implicit Bayes Adaptation: A Collaborative Transport Approach}
\name{Bo Jiang$^{\star}\thanks{Thanks to the generous support of ARO grant W911NF1910202.}$ \qquad Hamid Krim$^{\star}$ \qquad Tianfu Wu$^{\star}$ \qquad Derya Cansever$^{\dagger}$}
  
\address{$^{\star}$ North Carolina State University \\
    $^{\dagger}$US Army Research Office}

\begin{document}
\maketitle
\begin{abstract}
The power and flexibility of Optimal Transport  (OT) have pervaded a wide spectrum of problems, including recent Machine Learning challenges such as unsupervised domain adaptation. Its essence of quantitatively relating two probability distributions by some optimal metric, has been creatively exploited and shown to hold promise for many real-world data challenges. In a related theme in the present work, we posit that domain adaptation robustness is rooted in the intrinsic (latent) representations of the respective data, which are inherently lying in a non-linear submanifold embedded in a higher dimensional Euclidean space. We account for the geometric properties by refining the $l^2$ Euclidean metric to better reflect the geodesic distance between two distinct representations. We integrate a  metric correction term as well as a prior cluster structure in the source data of the OT-driven adaptation. 
We show that this is tantamount to an implicit Bayesian framework, which we demonstrate to be viable for a more robust and better-performing approach to domain adaptation. Substantiating experiments are also included for validation purposes.
\end{abstract}


\section{Introduction}
Much of the great success of Deep Learning is owed to the massive amounts of labeled data whose considerable effort of collection and  labeling is often overlooked\cite{jiang2022refining}. Its promise and viability largely ride on the flexibility of 
safeguarding the learning on a labeled data set to adapt to unlabeled data with possibly a different distribution. With a possibly significant semantic gap between  the first referred to as "source data", and the second, the "target data" (due to different sensing or other), an equivalence mapping is generally sought to not only exploit the extensive initial learning, but to also associate effective labeling whose space is assumed coincident with that of the source space.

Unsupervised domain adaptation essentially entails a minimization of the distance between the source and target domains using different criteria to reduce the distance, such as Maximum Mean Discrepancy (MMD)\cite{long2015learning}, Correlation Alignment\cite{sun2016deep}, and Kullback-Leibler  (KL) divergence\cite{zhuang2015supervised}. An adversarial learning approach was proposed in\cite{tzeng2017adversarial} to synthesize an adapted  equalizing mapping of the target domain to the source data, thereby facilitating the sought domain adaptation. Given its flexible amenability to assessing imbalance between distributions, optimal transport (OT) has also been central to reducing the distance between complex distributions\cite{damodaran2018deepjdot,xu2020reliable,fatras2021unbalanced,redko2017theoretical,patel2015visual}. Accounting for the complete statistical characteristics of both domains has remained absent, and using coarser domain knowledge together with a pair-wise association has fallen falling short in OT-based performance in domain adaptation. Similar samples, e.g., digit 1 and 7, will invariably result in gross mismatching in the transport plan. 
In computing the transport cost in practice, samples are often considered as a finite set of vectors in a high-dimensional Euclidean space. In many applications, however, these data points, also subjected to a likely non-linear representation\cite{jiang2021dynamicjournal}, will be distributed over a manifold rather than from a distribution with support over all of $\mathbb{R}^n$\cite{li2019geodesic,malik2018connecting,soliman2016mean}. A natural question to ask is how we can reconstruct or estimate the underlying manifold through some measured characteristics, e.g., curvatures. The challenge of adequately reconstructing the manifold to obtain a good geodesic estimate may yield low accuracy or high complexity with impediments to practical implementation.

To address the aforementioned issues in the existing approaches, we propose an OT-based cost formulation for domain adaptation. The proposed cost firstly helps preserve the underlying multi-class structure in each domain, where the true transport matrix should have a block-diagonal structure. In light of the intrinsicality of the latent representations to a submanifold embedded in a higher dimensional Euclidean space, our proposed alternative to a Euclidean distance is aimed at providing a refined estimate of the geodesic distance on the manifold. Following a discussion of some preliminary background on OT and domain adaptation in the next section, we develop our proposed approach in Section 3, where we first discuss a cluster-based OT and its associated adapted metric. Moreover, upon highlighting the importance of reducing the dependence of the captured latent embedding function on the classifier performance, we propose a dual collaborative OT framework which seeks to define a differential measure of goodness between an "agnostic" OT. In Section 4, we provide a detailed experimental assessment and evaluation of the proposed work, and some concluding remarks in Section 5.

\section{Background}
We adopt throughout the paper, $D$, $X$, and $Y$ to respectively denote the pixel space, the latent representation domain, and the label space. We use the lower case letter superscripts $s$ and $t$ to distinguish elements coming from the source or target, and the subscripted lower case letter to distinguish different elements in different spaces. We will also denote the cost matrix with the upper letter $C$, and the transportation as $\gamma$.  We let $f_\theta(\cdot)$ be a function parameterized by $W_\theta$.

\subsection{Unsupervised Domain Adaptation}
In an unsupervised domain adaptation setting, one may consider a set of source latent features $X^s=\{x^s_i\}_{i=1}^{N_s}$, which are extracted by a learnable embedding function $f_E(\cdot):D\rightarrow X$, with its corresponding set of labels $Y^s=\{y^s_i\}_{i=1}^{N_s}$, where $N_s$ is the total number of source samples. Let $f_C(\cdot):X\rightarrow Y$ be a classifier function that predicts the labels of the extracted latent representations.
Then, consider a set of unlabeled features from the target domain, $X^t=\{x^t_j\}_{j=1}^{N_t}$, where features may be generated by the same or additional embedding functions from the observed target data. While no label information is assumed known for the target data, our goal is to use the same classifier trained on $X^s$ to predict the labels of $X^t$, with an understanding that source and target data share the same label space. The predictions on target data would only achieve the best performance if $X^t$ has the same distribution as $X^s$.

Addressing this issue led to the idea of domain adaptation by minimizing the gap in the latent space. The domain adaptation problem is guaranteed by the following theorem,
\begin{theorem}
For any given hypothesis $h$ \cite{ben2010theory},\\
\begin{equation*}
\begin{split}
    \epsilon_t(h) & \leq \epsilon_s(h)+d(X^s,X^t)\\
    & +min\{E_{X^s}[|f_s(x)-f_t(x)|],E_{X^t}[|f_s(x)-f_t(x)|]\}
\end{split}
\end{equation*}
where $\epsilon(h)$ defines the error made by a hypothesis $h$ in the source or target domain, and $f_s(\cdot)$ and $f_t(\cdot)$ are true labelling function of source and target domain, respectively.
\end{theorem}
\noindent Theorem 1 shows that the target error of a given hypothesis is bounded by the summation of the source error, the distance between source and target domain, and an optimal difference.

\subsection{Optimal Transport}
Optimal transport (OT) theory was originally proposed in \cite{monge1781memoire} for the study of resource allocation problems. A geometric comparison of probability distributions is carried out by introducing a distance measure. The distance effectively transporting one distribution to another is achieved by minimizing the overall cost $C$. Formally speaking, let
$X^s$ and $X^t\in \mathbb{R}^n$ be two measurable spaces and denote the set of all probability measures over $X$ by $\mathcal{P}(X)$. Given two probability measures $\mu_s\in\mathcal{P}(X^s)$ and $\mu_t\in\mathcal{P}(X^t)$, the Monge-Kantorovich relaxation formulates an optimization of the transport cost, $\gamma^*=arg\min_\gamma\int_{X^s\times X^t}c(x^s,x^t)d\gamma(x^s,x^t)$, w.r.t. some cost measure $c(\cdot,\cdot): X^s\times X^t\rightarrow R_{0+}$, and the marginal constraints $\gamma(A\times X^t)=\mu_s(A), \gamma(X^s\times B)=\mu_t(B)$

In practice, with finite observations of distributions accessible to the experiments, we consider two discrete measures $\mu_s=\sum_{i=1}^{N_s}p_{x^s_i}\delta_{x^s_i}$ and $\mu_t=\sum_{j=1}^{N_t}p_{x^t_j}\delta_{x^t_j}$, where $\delta_x$ represents the Dirac function at point $x$, and $\sum_{i=1}^{N_s}p_{x^s_i}=\sum_{j=1}^{N_t}p_{x^t_j}=1,~p_{x^s_i},p_{x^t_j}\geq0$. The optimal transport in the discrete case is defined as following:
\begin{equation}
\begin{split}
    & \gamma^*=arg\min_{\gamma\in\Gamma(\mu_s,\mu_t)}\langle C,\gamma\rangle=\sum_{i=1}^{N_s}\sum_{j=1}^{N_t}\gamma_{ij}C_{ij},\\
    & \Gamma(\mu_s,\mu_t)=\{\gamma\in R_{0+}^{N_s\times N_t}|\gamma\mathbf{1}_{N_s}=\mu_s,\gamma^T\mathbf{1}_{N_s}=\mu_t\},
\end{split}
\end{equation}
where $C$ is the cost matrix computed by a pairwise cost measure $C_{ij}=c(x^s_i,x^t_j)$,  with $L_2$-norm as a commonly used cost function, and $\mathbf{1}_n$ is a vector of ones. Note that the discrete version of optimal transport is a linear programming task, with a dual formulation and alternative solutions exploited in various recent applications.

\subsection{Geodesic Distance}
\begin{figure}
  \centering
  \includegraphics[scale = 0.5]{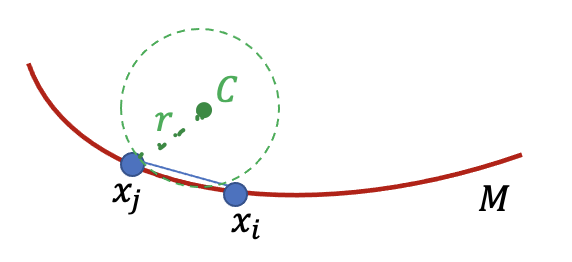}
  \caption{Comparison between Euclidean and Geodesic distance.}
  \label{fig:distance_manifold}
\end{figure}

Most machine learning approaches rely on pairwise distances between samples, and the choice of distance metric has a significant impact on the performance\cite{li2019geodesic}. The selection of such a metric depends on the data and the associated representation. As illustrated in Fig.\ref{fig:distance_manifold}, the widely applied Euclidean distance, hence, is suited for when the data are concentrated on an unknown submanifold, and the intrinsic geometry of the data needs to be taken into account. An associated geodesic is the shortest path along the manifold with its geometric properties.

Recall that a geodesic is a curve on a manifold whose velocity is constant\cite{lee2009manifolds}. For a given $M$ be a differentiable manifold, let $x\in M$, and let $v\in T_xM$ be a tangent vector to the manifold at $x$. Then, a unique geodesic $\gamma_v$ satisfies, $\gamma_v(0)=x$ and $\gamma_v^'(0)=v$.
The corresponding exponential map is defined by $exp_x(v)=\gamma_v(1)$ locally in general. Moreover, Hopf–Rinow theorem\cite{hopf1931ueber, atkin1975hopf} asserts that in the case of a complete finite-dimension manifold as a metric space, the exponential map is defined on the whole tangent space.

We propose in the next section, an alternative approach  to the domain adaptation problem that yields an improved solution when co-opted with the proper geometry of the related ambient data spaces.

\section{Equalizing Domain Variation: An Optimal Transport Approach}
In introducing here, our proposed model first invoke the geometric structure of the conventional cost matrix, and relax its evaluation using a near-geodesic metric. By next adopting a Bayesian approach when invoking some problem-associated parameters, we introduce a collaborative optimal transport learning, with a proper geodesic-based cost together with the underlying structures of the predicted and true label information. The double-map transport learning collaboratively exploits the information from both predicted and true labels, and reduces the effect of the uncertainty of the training classifier.

\subsection{Cluster-to-Cluster Optimal Transport}
As noted in the previous section, one assumption in the existing approaches is that all data points lie in a Euclidean space, hence ignoring the underlying intrinsic space of the latent data. Bearing in mind that DA is fundamentally an inference problem, which entails the minimization of the intra-class distance between entities and the maximization of their inter-class distance. This class-induced clustering structure of the source and target domains injects some constraining ordering property which we implicitly introduce as a Bayesian prior.
The source data labels are accessible as $X^s=\{X^s_k\}_{k=1}^K$, where $K$ is the total number of classes. The predictions of target labels $\hat{y}^t_j$ can be assigned by using the classifier or applying the $k$-nearest-neighbor ($k$-NN) classification method as, $X^t=\{X^t_k\}_{k=1}^K$. The cluster-to-cluster optimal transport (CCOT) is defined as following:
\begin{equation}
    CCOT(X^s,X^t)=\sum_kOT(X^s_k,X^t_k),
\end{equation}
Using the label information as a basis for the source domain structure, yields an associated point cloud with a submanifold ambient space structure embedded in a $n$-dimensional Euclidean space. The intrinsic geometry thus imposes a modification of the pairwise transport cost given in Eq.(2). The geometric influence on the cost matrix is analytically quantified as follows,
\begin{proposition}
The following statements are equivalent:\\
(a) The optimal coupling between corresponding clusters is $\gamma_k^*=\min_{\gamma_k}\langle C_k,\gamma_k\rangle$, where $C_{k,ij}=\|x^s_i-x^t_j\|_2^2$ and $x^s_i,x^t_j$ are in $k^{th}$ class.\\
(b) When source and target domains have not been reordered,\\ $\hat{C}_{ij}=
    \begin{cases}
      \|x^s_i-x^t_j\|^2_2, & \text{if}\ y^s_i=\hat{y}^t_j \\
      \infty, & \text{otherwise}
    \end{cases}$.
\end{proposition}

\subsection{Relaxation of CCOT}
To account for the geometric properties of the relevant data space, the cost matrix is properly adapted to  the cluster-to-cluster optimal transport, as just noted. This approximate refinement imposes, however, a constraint on the samples from different classes, preventing a prediction to cross a decision boundary. In other words, the transportation is only true when the information of target labels is known and accessible. We consequently, relax the infinite cost between inter-class connections by a penalty to preserve the block-diagonal structure of the transport matrix, $C_{ij}=\|x^s_i-x^t_j\|_2^2+m$, if $y^s_i\neq \hat{y}^t_j$. A straightforward and naive implementation, however, yields a discontinuity in the cost function on account of the hard-decision condition, thus causing a jump for samples near and on different sides  the decision boundary. This discontinuity in the optimization will cause Dirac delta singularity in the gradient calculation, making it non-smooth, and affecting the convergence with induced oscillations in the optimization.
To mitigate this difficulty,we proceed to (1) Soften the hard decision boundary allowing for cross-class decisions; (2) Safeguard the geometric properties underlying the space and reconstruct associated geodesic metric.

On account of exponential map's analytical expression, we may view the label space as a tangent space on the manifold, and this newly proposed pairwise transport cost can thus be written as,
\begin{equation}
    C_{ij}=c(x^s_i,x^t_j)=\alpha\|x^s_i-x^t_j\|^2_2+(1-\alpha)e^{\beta\|\tilde{y}^s_i-\tilde{y}^t_j\|^2_2},
\end{equation}
where $\tilde{y}$ is some label information of latent representations, $\alpha$ is the convex combination parameter that balances the Euclidean distance and the geodesic-induced recovery information, $\beta$ controls the geodesic velocity on the tangent space.

\begin{proposition}
The optimal transport between the true source label ${\tilde Y}^s$ and $k$-NN -based estimated target label ${\tilde Y}^t$, is based on an implicit prior $p(k)$ to achieve the optimal cluster-to-cluster mapping.
\end{proposition}
\noindent Proposition 2 highlights the implicit prior knowledge of class probability contained in the modified cost matrix. Moreover, it is the smoothness to the transportation between the clusters of the observed data that have the highest joint probability.

\subsection{Collaborative Optimal Transport}
As shown in Eq.(3), the proposed geodesic distance aims to mollify the cost with respect to the label information of both source and target domains under any possible classification and subsequent transport/assignment and to mitigate a number of limitations. At first, the adopted classifier, on the one hand, will highly influence the recovery quality of the manifold structure, and its uncertainty, on the other hand, will impact the embedding function (e.g., only reaching a local minimal during training). The estimated source label in computing the discrepancy between source and target domains will, in addition, affect the clustering membership information in the model. To help alleviate this difficulty, we propose an implicit Bayesian approach using prior information in carrying out a (more informed) transport  to be contrasted with its fully agnostic counterpart, and use the deviation between the two for a systematic adaptation. This so-referred collaborative transport  as we next elaborate, helps in better accounting for the underlying manifold structure of the latent data, and in improving the training for the embedding function.

It is clear that the inclusion of an additional classifier can effectively decrease the perturbation (error) effect of the training classifier and increase exchangeable clustering information from different classification methods. We proceed to seek two transport maps that are based on different label information, using a convex combination balancing the information introduced by the two classifiers. The distance between source and target domains via
collaborative optimal transport can thus be written as:

\begin{equation}
    \mathcal{L}_{OT}=\min_{\gamma_1,\gamma_2}\alpha\langle\gamma_1,\Tilde{C}_1\rangle+(1-\alpha)\langle\gamma_2,\Tilde{C}_2\rangle,
\end{equation}
where the elements of $\Tilde{C}_1$ and $\Tilde{C}_2$ are calculated via Eq.(4) with predicted labels, $\hat{Y}^s$ and $\hat{Y}^t$, given by the classifier $f_C(\cdot)$, and true source labels $Y^s$ and alternative predicted target labels $\hat{Y}^{t'}$, given by the $k$-NN method, respectively. 
We want the embedding function to collaboratively take advantage of information from both sides, and thus the two optimal transport maps should be further constrained tightly due to the fact of transportation between the same pair of distributions, i.e., source and target domains.

\begin{proposition}
A dually collaborative transport map pair $\gamma_1$ and $\gamma_2$, converges to a single transport map at training completion, yielding a vanishing distance between the target predicted labels, $\hat{y}^t_j$ and $\hat{y}^{t'}_j$, given by different classifiers, with a proportional accuracy improvement over that of a single classifier.
\end{proposition}
\noindent\textbf{Proof:} At the conclusion of sufficient training, any classifier error between the predicted source labels $\hat{Y}^s$ and the true source labels $Y^s$ is sought to be $\|\hat{y}^s_i-y^s_i\|^2_2\leq\epsilon$, where $\epsilon$ is some small positive value, and $\hat{y}^s_i=f_C(x^s_i)$. Given the optimal transport map as a joint distribution, showing the probability of one source sample being transported to one target sample, the modified cost reveals the additional information inserted the original distance expression. The corresponding optimal transportation is thus the conditional joint probability, where predicted and true labels support the conditions, $\gamma^*_{1,ij}\coloneqq p(x^s_i,x^t_j|\hat{y}^s_i,\hat{y}^t_j,\Theta)$ and $\gamma^*_{2,ij}\coloneqq p(x^s_i,x^t_j|y^s_i,\hat{y}^{t'}_j,\Theta)$, where $\Theta$ is a set of model parameters and the optimal transport constraints. $Y$ is deterministic given a defined classification function, the term $p(y|x,\Theta)$ is then determined. Based on Bayes' rule and the convergence of $\hat{y}^s_i$ to $y^s_i$ by the supervised learning of source data, and the modified cost is relaxed to a fully smooth formula in terms of classes and decision bounds, the convergence between two transport maps infers that $p(x^s_i,x^t_j|\hat{y}^s_i,\hat{y}^t_j,\Theta)$ and $p(x^s_i,x^t_j|y^s_i,\hat{y}^{t'}_j,\Theta)$ should converge to the same point., which leads to $p(\hat{y}^t_j)=p(\hat{y}^{t'}_j)$. The minimization of gaps between two transportation plans collaborates the local information by different classifiers, which highly reduces the prediction uncertainty by a single classifier.  \hfill$\blacksquare$

The idea of the proposed model is to further shrink the search space for the optimal transport map by constraining the map to satisfy  both classification conditions. The convergence of the two transport maps to one single map with respect to Proposition 3, is achieved by minimizing the Kullback–Leibler (KL) divergence\cite{kullback1951information}.

\subsection{Algorithmic Solution}
As noted in Theorem 1, the target-error bound also depends on the source error. To attain the optimal latent model of the source data, we can learn the embedding function $f_E(\cdot)$ and the classifier $f_C(\cdot)$ by minimizing the cross-entropy loss between the ground truth labels $Y^s$ and the predictions $\hat{Y}^s=f_C(f_E(X^t))$, $\min_{f_E,f_C}\mathcal{L}_{CE}(Y^s,\hat{Y}^s)$, where $\mathcal{L}_{CE}$ denotes the cross-entropy loss.

In \cite{cuturi2013sinkhorn} an entropic regularization term was proposed to reduce  the search space, and to achieve a computationally more efficient near-optimal transport. Similarly, a regularization of the predicted  target samples is desired even if inaccessible to the training. It reduces the candidate space and also favors a model with more confident predictions of target samples, with a minimum entropy value vector achieved by a one-hot vector, which helps preserve distinct classes and maximize the inter-class "distance" in the target domain.

To sum up, the objective loss is written as:
\begin{equation}
    \mathcal{L}=\mathcal{L}_{CE}(Y^s,\hat{Y}^s)+\lambda_1\mathcal{L}_{OT}+\lambda_2\mathcal{H}(\hat{Y}^t)+\lambda_3KL(\gamma^*_1||\gamma^*_2),
\end{equation}
where $\mathcal{L}_{OT}$ is the collaborative optimal transport loss defined in Eq.(4), $\gamma^*_i$, $i=1,2$, are the optimal transport plans based on the cost matrix $\Tilde{C}_i$, the elements of $\Tilde{C}_i$ is defined by Eq.(3) with different label information. To effectively account for the afore-discussed class Bayes prior using the underlying true class label for the second collaborative transport comparative,  we apply a
$k$-NN algorithm on $\left(X^s, Y^s, X^t\right)$ to achieve alternative target label information. A numerically enhanced and stable clustering estimator is secured by way of a memory bank \cite{he2020momentum} on the source data. $\mathcal{H}(a)=-\sum_ia_i\log a_i$ is the entropic regularization of each target prediction, $KL(A||B)=-\sum_{i,j}A_{ij}\log B_{ij}$, and $\lambda_1$, $\lambda_2$, and $\lambda_3$ are three hyper-parameters, balancing the contribution to the total loss.

\section{Experiments and Results}
To evaluate our proposed approach, we use some widely accepted domain adaptation datasets, for validation purposes as well as for comparison to several state-of-the art methods with available online codes \cite{tzeng2017adversarial,damodaran2018deepjdot,roheda2022fast}.

\subsection{Datasets}
The first task is adaptation three standard digit classification datasets: MNIST, USPS, and SVHN. Each dataset consists of 10 classes of digits, from 0 to 9. USPS is a low resolution dataset, while SVHN is more complex as each sample image is a mixed set of numbers in the same image. We follow existing works to construct three transfer tasks: USPS$\rightarrow$MNIST, MNIST$\rightarrow$USPS and SVHN$\rightarrow$MNIST.

Office-31 is an advanced domain adaptation dataset. It consists of 31 categories in three domains: Amazon (W), DSLR (D), and Webcam (W). The 31 categories in the dataset include of common objects in an office area, such as keyboards, calculator, and headphones.  By permuting the three domains, we obtain following adaptation tasks: A$\rightarrow$W, A$\rightarrow$D, D$\rightarrow$W, W$\rightarrow$D, D$\rightarrow$A, and W$\rightarrow$A.

\begin{figure}[h]
  \centering
  \includegraphics[width = 6.5cm, height = 3.2cm]{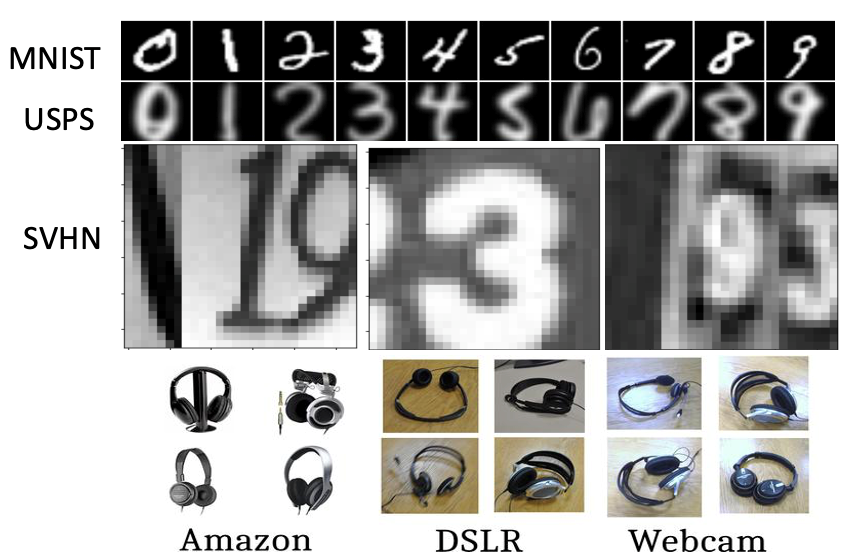}
  \caption{Examples of datasets.}
  \label{fig:samples}
\end{figure}

\subsection{Experiments and Results}
In our experiments, different architectures for the embedding networks were used fin different data set scenarios:
All digits were processed using LeNet-5;
Office-31 data used Res-Net-50\cite{he2016deep} as the backbone network.

Each model\cite{tzeng2017adversarial,damodaran2018deepjdot,roheda2022fast} started with a pre-trained neural network which was  further fine-tuned accordingly. In all considered cases, a linear classifier (a single-layer network) is trained simultaneously with the embedding function. The final prediction and accuracy of the target dataset are based on this trained linear classifier.
The neural network is optimized by the Stochastic Gradient Descent (SGD) optimizer with a momentum of 0.9 and a batch size of 128. The learning rate is constant over the training procedure. All hyper-parameters are carefully tuned to yield the best results.

Tables 1 and 2 list the performance of our proposed model compared to a series of other domain adaptation methods on Digit, and Office-31, respectively. As can be seen, the proposed model markedly outperforms all other models on all tasks. Moreover, Our model shows better stability and improvement for both simple and more complex tasks in comparison with existing OT-based approaches. The label information provided by different classifiers shrinks the search space of the optimal transport, and stabilizes the information adaptation between different domains. Moreover, the two collaborative classifiers will also prevent each other from being stuck at some local minimum during the training procedure, due to the convergence between predicted labels. 
\begin{table}[h]
\footnotesize
\renewcommand{\arraystretch}{1}
    \centering
    \begin{tabular}{|c||c|c|c|}
    \hline
    Method & MNIST$\rightarrow$USPS & USPS$\rightarrow$MNIST & SVHN$\rightarrow$MNIST\\
    \hline
    ADDA & 93.5 & 94.3 & 86.2\\
    \hline
    Fast OT & 96.6 & 94.6& 88.4\\
    \hline
    DeepJDOT & 96.3 & 95.8 & 92.7\\
    \hline
    Our model & 97.1 & 96.5 & 93.1\\
    \hline
    \end{tabular}
    \caption{Digit number results}
    \label{tab:digit}
\end{table}

\begin{table}[h]
\renewcommand{\arraystretch}{1}
\footnotesize
    \centering
    \begin{tabular}{|c||c|c|c|c|c|c|}
    \hline
    Method & A$\rightarrow$W & A$\rightarrow$D & D$\rightarrow$W & W$\rightarrow$D & D$\rightarrow$A & W$\rightarrow$A\\
    \hline
    ADDA & 81.2 & 78.2 & 92.8 & 99.1 & 63.2 & 64.5\\
    \hline
    Fast OT & 83.9 & 82.2 & 93.2 & 99.0 & 65.2 & 66.1\\
    \hline
    DeepJDOT & 86.2 & 85.5 & 95.2 & 99.6 & 67.1 & 69.1\\
    \hline
    Our model & 87.3 & 86.2 & 95.8 & 99.6 & 71.5 & 72.1\\
    \hline
    \end{tabular}
    \caption{Office-31 results}
    \label{tab:office_31}
\end{table}

\section{Conclusions}
In this paper, we proposed a collaborative learning model for unsupervised domain adaptation using optimal transport theory and geodesic distance recovery. The proposed method first aims at constructing a meaningful manifold, which includes clustering information, for both source and target domains. We propose a modified cost to recover the geodesic distance on the manifold based on the exponential map of Riemannian geometry. The structural information is also contained and relaxed by modifying the cost matrix with the information of true and predicted labels. We then propose a dual-map collaborative-learning system to exploit information across contrasting  classifiers to minimize their discrepancy. The proposed model first builds up an inner-domain structure, and then compute transportation between corresponding clusters to reduce mismatching between samples near the decision boundary. Finally, our proposed model provided improved adaptation performance on numerous domain datasets.

\newpage
\bibliographystyle{unsrt}
{{\footnotesize \bibliography{ref.bib}}

\begin{thebibliography}{10}

\bibitem{jiang2022refining}
Bo~Jiang, Hamid Krim, Tianfu Wu, and Derya Cansever.
\newblock Refining self-supervised learning in imaging: Beyond linear metric.
\newblock {\em arXiv preprint arXiv:2202.12921}, 2022.

\bibitem{long2015learning}
Mingsheng Long, Yue Cao, Jianmin Wang, and Michael Jordan.
\newblock Learning transferable features with deep adaptation networks.
\newblock In {\em International conference on machine learning}, pages 97--105.
  PMLR, 2015.

\bibitem{sun2016deep}
Baochen Sun and Kate Saenko.
\newblock Deep coral: Correlation alignment for deep domain adaptation.
\newblock In {\em European conference on computer vision}, pages 443--450.
  Springer, 2016.

\bibitem{zhuang2015supervised}
Fuzhen Zhuang, Xiaohu Cheng, Ping Luo, Sinno~Jialin Pan, and Qing He.
\newblock Supervised representation learning: Transfer learning with deep
  autoencoders.
\newblock In {\em Twenty-Fourth International Joint Conference on Artificial
  Intelligence}, 2015.

\bibitem{tzeng2017adversarial}
Eric Tzeng, Judy Hoffman, Kate Saenko, and Trevor Darrell.
\newblock Adversarial discriminative domain adaptation.
\newblock In {\em Proceedings of the IEEE conference on computer vision and
  pattern recognition}, pages 7167--7176, 2017.

\bibitem{damodaran2018deepjdot}
Bharath~Bhushan Damodaran, Benjamin Kellenberger, R{\'e}mi Flamary, Devis Tuia,
  and Nicolas Courty.
\newblock Deepjdot: Deep joint distribution optimal transport for unsupervised
  domain adaptation.
\newblock In {\em Proceedings of the European Conference on Computer Vision
  (ECCV)}, pages 447--463, 2018.

\bibitem{xu2020reliable}
Renjun Xu, Pelen Liu, Liyan Wang, Chao Chen, and Jindong Wang.
\newblock Reliable weighted optimal transport for unsupervised domain
  adaptation.
\newblock In {\em Proceedings of the IEEE/CVF Conference on Computer Vision and
  Pattern Recognition}, pages 4394--4403, 2020.

\bibitem{fatras2021unbalanced}
Kilian Fatras, Thibault S{\'e}journ{\'e}, R{\'e}mi Flamary, and Nicolas Courty.
\newblock Unbalanced minibatch optimal transport; applications to domain
  adaptation.
\newblock In {\em International Conference on Machine Learning}, pages
  3186--3197. PMLR, 2021.

\bibitem{redko2017theoretical}
Ievgen Redko, Amaury Habrard, and Marc Sebban.
\newblock Theoretical analysis of domain adaptation with optimal transport.
\newblock In {\em Joint European Conference on Machine Learning and Knowledge
  Discovery in Databases}, pages 737--753. Springer, 2017.

\bibitem{patel2015visual}
Vishal~M Patel, Raghuraman Gopalan, Ruonan Li, and Rama Chellappa.
\newblock Visual domain adaptation: A survey of recent advances.
\newblock {\em IEEE signal processing magazine}, 32(3):53--69, 2015.

\bibitem{jiang2021dynamicjournal}
Bo~Jiang, Yuming Huang, Ashkan Panahi, Yiyi Yu, Hamid Krim, and Spencer~L
  Smith.
\newblock Dynamic graph learning: A structure-driven approach.
\newblock {\em Mathematics}, 9(2):168, 2021.

\bibitem{li2019geodesic}
Didong Li and David~B Dunson.
\newblock Geodesic distance estimation with spherelets.
\newblock {\em arXiv preprint arXiv:1907.00296}, 2019.

\bibitem{malik2018connecting}
John Malik, Chao Shen, Hau-Tieng Wu, and Nan Wu.
\newblock Connecting dots--from local covariance to empirical intrinsic
  geometry and locally linear embedding.
\newblock {\em arXiv preprint arXiv:1804.02811}, 2018.

\bibitem{soliman2016mean}
YOUSUF~M SOLIMAN.
\newblock Mean and principal curvature estimation from noisy point cloud data
  of manifolds embedded in rn.
\newblock 2016.

\bibitem{ben2010theory}
Shai Ben-David, John Blitzer, Koby Crammer, Alex Kulesza, Fernando Pereira, and
  Jennifer~Wortman Vaughan.
\newblock A theory of learning from different domains.
\newblock {\em Machine learning}, 79(1):151--175, 2010.

\bibitem{monge1781memoire}
Gaspard Monge.
\newblock M{\'e}moire sur la th{\'e}orie des d{\'e}blais et des remblais.
\newblock {\em Histoire de l'Acad{\'e}mie Royale des Sciences de Paris}, 1781.

\bibitem{lee2009manifolds}
Jeffrey~M Lee, Bennett Chow, Sun-Chin Chu, David Glickenstein, Christine
  Guenther, James Isenberg, Tom Ivey, Dan Knopf, Peng Lu, Feng Luo, et~al.
\newblock Manifolds and differential geometry.
\newblock {\em Topology}, 643:658, 2009.

\bibitem{hopf1931ueber}
Heinz Hopf and Willi Rinow.
\newblock Ueber den begriff der vollst{\"a}ndigen differentialgeometrischen
  fl{\"a}che.
\newblock {\em Commentarii Mathematici Helvetici}, 3:209--225, 1931.

\bibitem{atkin1975hopf}
Christopher~J Atkin.
\newblock The hopf-rinow theorem is false in infinite dimensions.
\newblock {\em Bulletin of the London Mathematical Society}, 7(3):261--266,
  1975.

\bibitem{kullback1951information}
Solomon Kullback and Richard~A Leibler.
\newblock On information and sufficiency.
\newblock {\em The annals of mathematical statistics}, 22(1):79--86, 1951.

\bibitem{cuturi2013sinkhorn}
Marco Cuturi.
\newblock Sinkhorn distances: Lightspeed computation of optimal transport.
\newblock {\em Advances in neural information processing systems}, 26, 2013.

\bibitem{he2020momentum}
Kaiming He, Haoqi Fan, Yuxin Wu, Saining Xie, and Ross Girshick.
\newblock Momentum contrast for unsupervised visual representation learning.
\newblock In {\em Proceedings of the IEEE/CVF conference on computer vision and
  pattern recognition}, pages 9729--9738, 2020.

\bibitem{roheda2022fast}
Siddharth Roheda, Ashkan Panahi, and Hamid Krim.
\newblock Fast ot for latent domain adaptation.
\newblock {\em arXiv preprint arXiv:2210.00479}, 2022.

\bibitem{he2016deep}
Kaiming He, Xiangyu Zhang, Shaoqing Ren, and Jian Sun.
\newblock Deep residual learning for image recognition.
\newblock In {\em Proceedings of the IEEE conference on computer vision and
  pattern recognition}, pages 770--778, 2016.

\end{thebibliography}

\end{document}